# No "Evidence for a new phase of dense hydrogen above 325 GPa"

Ranga P. Dias, Ori Noked, and Isaac F. Silvera

Lyman Laboratory of Physics, Harvard University, Cambridge MA, 02138

In recent years there has been intense experimental activity to observe solid metallic hydrogen. Wigner and Huntington predicted that under extreme pressures insulating molecular hydrogen would dissociate and transition to atomic metallic hydrogen. Recently Dalladay-Simpson, Howie, and Gregoryanz reported a phase transition to an insulating phase in molecular hydrogen at a pressure of 325 GPa and 300 K. Because of its scientific importance we have scrutinized their experimental evidence to determine if their claim is justified. Based on our analysis, we conclude that they have misinterpreted their data: there is no evidence for a phase transition at 325 GPa.

The purpose of this paper is to carefully re-examine and critically analyze the recently published paper in Nature by Dalladay-Simpson, Howie, and Gregoryanz (**DHG**) [1], entitled *"Evidence for a new phase of dense hydrogen above 325 GPa"*. In their paper, DHG claim to observe a new phase at a pressure of 325 GPa at room temperature, which they named phase V. This claim is based on observed changes in the measured Raman modes. Subsequently, Eremets, Troyan, and Drozdov (**ETD**) [2] presented new Raman data in a similar pressure regime and at room temperature, which do not support the claim made by DHG. ETD show that phases IV and V have a large pressure range (270-320 GPa) in which the two phases coexist and the hydrogen sample becomes pure phase V only at higher pressures. ETD give several plausible reasons that could explain the observations of DHG as being a misinterpretation of data for phase IV' [3-7], as a mixed state of phase IV and phase V.

Ideally the structures of phases are determined by x-ray or neutron diffraction techniques, but these are very challenging measurements for the hydrogens. The weak scattering cross sections and the very small sample size render conventional diffraction techniques ineffectual at extreme pressures. The main technique for the study of the high-pressure phases of hydrogen has been lattice vibrational spectroscopy, i.e. Raman scattering and IR absorption. According to the Landau theory of phase transitions (**PT**), based on symmetry of the particle distributions, a PT is characterized by an order-parameter that is continuous for a second-order PT and discontinuous for a first-order PT. Group theoretical arguments show that excitations (vibrons, rotons, phonons, etc.) are allowed or disallowed, depending on the symmetry of the structure. Thus, at a phase transition, modes appear, disappear, or undergo discrete shifts as the symmetry of the lattice changes, and these changes are used to map the phase lines.

Now we focus our attention on the claim by DHG. The authors use the following criteria for detecting a PT in hydrogen:

1. Substantial decrease in intensity of the high frequency vibron modes

2. Changes in position, width and intensity of the low frequency lattice modes

3. Disappearance of low frequency lattice modes

4. Change of the high frequency vibron mode's frequency vs. pressure slope

According to conventional criteria, only No. 3 above is valid.

The first criterion used by DHG claims that a substantial decrease in intensity of the high frequency vibron modes can be observed. In Fig. 1 we re-plot the pressure dependence of the relative intensity of the $\upsilon_1$ vibron mode given in Fig. 2 of DHG. This plot of intensity vs. pressure shows a gradual change in intensity that, to within the uncertainties, can be fit to a linear function, with no abrupt change at 325 GPa that might accompany a PT. The $\upsilon_1$ mode has a strong overlap with the second-order diamond line that spans the 2300 -2600 cm$^{-1}$ frequency range. It is not clear how the intensity of the vibron mode is extracted in this region, as the second-order diamond line (see their Extended data, Fig. 7) also has a pressure dependence. The finite variation of mode intensity is not a signature of a PT. On the other hand, this behavior is what one might expect for a coexisting phase region with one phase dying out with pressure, as pointed out by ETD.

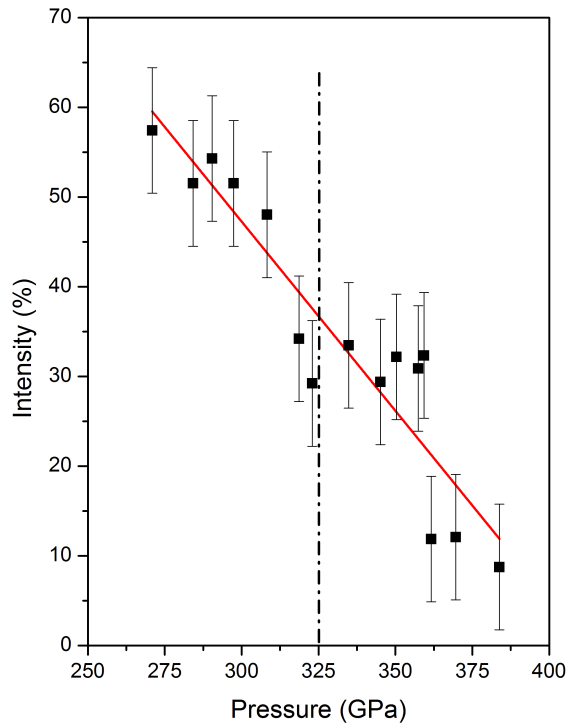

Fig. 1. The pressure dependence of the relative intensity of $\upsilon_1$ mode observed by DHG from their Fig. 2. The red line is a linear fit to all points; separate linear fits for the range 250-325 GPa and 325-400 GPa, show an insignificant change in slope.

The second criterion of DHG is that a substantial change in the Raman line width of the $L_1$ mode of hydrogen is indicative of a phase transition. DHG have claimed a rapid increase in the full-width-half-maximum (FWHM) above 325 GPa. Figure 2 shows the pressure dependence of the FWHM of the $L_1$ mode, from Fig. 2 of DHG. For clarity we have colored all of their data points for the $L_1$ mode of hydrogen red. No change is seen around 325 GPa, while a change is observed starting at around 360 GPa. Thus, the DHG claim of a PT at 325 GPa, based in part on a change of the $L_1$ line width is not supported by their data.

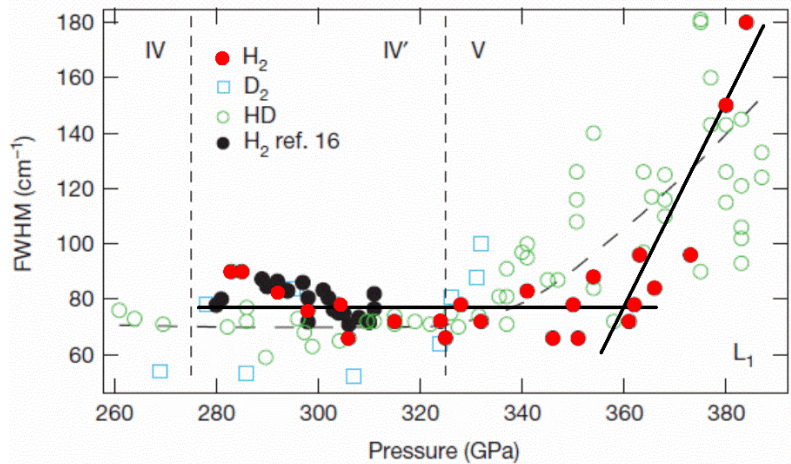

Fig. 2. The pressure dependence of the FWHM of the $L_1$ modes observed by DHM. The red filled circles indicate the FWHM for the hydrogen $L_1$ mode. The solid black lines are guides to the eye.

The third criterion claims that the data exhibit a clear disappearance of low frequency lattice modes. Disappearing or appearing of new modes is a definitive indication of a phase transition. DHG claim that two low frequency lattice modes, the $L_2$ and $L_3$ modes disappear at 325 GPa. The $L_2$ mode is clearly present in their Fig. 1 to at least 350 GPa and they even plot the pressure dependence in their Fig. 3c, which we reproduce in our Fig. 3. DHG also reported disappearance of the $L_3$ mode; this can be explained by masking of the mode due to overlap with the strong broad first-order diamond phonon line that starts at 1332 cm$^{-1}$ (see their Fig. 1a or Extended data Fig. 1). In Fig. 3 we re-plot the pressure dependence of the peak frequencies of the low frequency modes shown in Fig. 3c of DHG. The blue dash-dot line indicates the diamond phonon position at 1332 cm$^{-1}$ and the vertical brown dash-dot line indicates the pressure around 320 GPa where $L_3$ moves under the diamond line. At 325 GPa the L3 mode clearly overlaps the diamond line; thus, the mode is hidden rather than disappearing. The black dash-dot line indicates the region where phase the coexistence of phases IV and V begins. Therefore, while the criterion of disappearing modes for a phase transition is valid, the evidence presented by DHG for the claimed phase transition is not convincing.

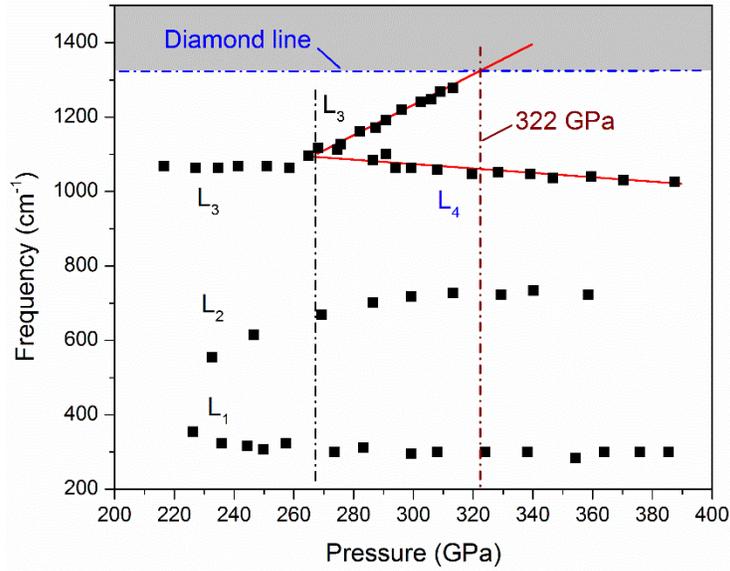

Fig. 3. The pressure dependence of the low frequency modes observed by DHG. The blue dash-dot line indicates the broad diamond phonon position beginning at 1332 cm$^{-1}$; the vertical brown dash-dot line indicates the pressure around 320 GPa where L3 moves under the diamond line. At 325 GPa the $L_3$ mode clearly overlaps the diamond line and the mode is hidden rather than disappearing. The black dash-dot line indicates the region where phase V begins to mix with phase IV. We also note the continuity of the $L_1$ and $L_2$ modes beyond this line.

The fourth criterion claims that the change of slope of the $\upsilon_1$ mode frequency vs pressure plot is an indication of phase transition. There are two difficulties to this specific claim. First, in Fig. 1a of DHG it is clear that the $\upsilon_1$ mode strongly overlaps the second-order diamond line and thus, similar to the critique of the second criterion above, the mode frequencies cannot be accurately determined. Fig. 3c of DHG does not show experimental uncertainties in the mode frequency that are important to support their claim. Second, although a slope may change at a PT, a change of slope is not a signature of a PT, which generally involves a change in symmetry or an order parameter. ETD also argue that the change of slope is erroneous.

In conclusion it is our opinion that the new data presented by Dalladay-Simpson, Howie, and Gregoryanz provide no evidence for a new phase transition in dense hydrogen at 325 GPa. Even if their criteria for a phase transition were acceptable, not one of their four criteria withstands rigorous scrutiny. Phase IV', between 270 GPa and 320 GPa, should be interpreted as a coexistence of phase IV and Phase V. Phase V has been documented in literature by many groups in this field [2].